\documentclass[draft]{agujournal2019}
\usepackage{url} 
\usepackage{lineno}
\usepackage{soul}
\usepackage{lscape}
\usepackage{rotating}
\usepackage{makecell}
\usepackage[utf8]{inputenc}
\usepackage[T1]{fontenc}
\usepackage{lmodern}
\usepackage{graphicx}
\usepackage{url}
\usepackage{setspace}
\usepackage[figurename=Fig.,labelfont=bf,labelsep=period]{caption}
\usepackage{subcaption}
\usepackage{amsmath}
\usepackage{xspace}
\usepackage{microtype}

\newcommand{\Hydrograd}{\textit{Hydrograd}\xspace}

\draftfalse

\journalname{}

\begin{document}

\title{Scientific Machine Learning of Flow Resistance Using Universal Shallow Water Equations with Differentiable Programming}

\authors{Xiaofeng Liu\affil{1,2} and Yalan Song\affil{1}}

\affiliation{1}{Department of Civil and Environmental Engineering, Pennsylvania State University, University Park, Pennsylvania, USA 16802.}
\affiliation{2}{Institute of Computational and Data Sciences, Pennsylvania State University, University Park, Pennsylvania, USA 16802.}

\correspondingauthor{Xiaofeng Liu}{xzl123@psu.edu}
\correspondingauthor{Yalan Song}{yxs275@psu.edu}

\begin{keypoints}
\item We developed a Universal Shallow Water Equations (USWEs) solver using differentiable programming for gradient tracking and seamless integration with neural networks.

\item We successfully used USWE to inversely model the spatial distribution of Manning’s $n$ within a real-world river channel.

\item With the support of a neural network, we recovered a universal relationship between flow resistance, water depth, and flow velocity across the channel.

\end{keypoints}

\begin{abstract}
Shallow water equations (SWEs) are the backbone of most hydrodynamics models for flood prediction, river engineering, and many other water resources applications. 
The estimation of flow resistance, i.e., the Manning's roughness coefficient $n$, is crucial for ensuring model accuracy, and has been previously determined using empirical formulas or tables. To better account for temporal and spatial variability in channel roughness, inverse modeling of $n$ using observed flow data is more reliable and adaptable; however, it is challenging when using traditional SWE solvers. Based on the concept of universal differential equation (UDE), which combines physics-based differential equations with neural networks (NNs), we developed a universal SWEs (USWEs) solver, \Hydrograd, for hybrid hydrodynamics modeling. It can do accurate forward simulations, support automatic differentiation (AD) for gradient-based sensitivity analysis and parameter inversion, and perform scientific machine learning for physics discovery. In this work, we first validated the accuracy of its forward modeling, then applied a real-world case to demonstrate the ability of USWEs to capture model sensitivity (gradients) and perform inverse modeling of Manning's $n$. Furthermore, we used a NN to learn a universal relationship between $n$, hydraulic parameters, and flow in a real river channel. Unlike inverse modeling using surrogate models, \Hydrograd uses a two-dimensional SWEs solver as its physics backbone, which eliminates the need for data-intensive pretraining and resolves the generalization problem when applied to out-of-sample scenarios. This differentiable modeling approach, with seamless integration with NNs, provides a new pathway for solving complex inverse problems and discovering new physics in hydrodynamics. 
\end{abstract}

\section{Introduction}\label{sec:introduction}

Hydrodynamic models have been used widely in the fields of river engineering, geomorphology, environmental science, and water resources engineering. Shallow water equations (SWEs) solvers are important hydrodynamic models for various applications in hydrology and hydraulics. Within SWEs, flow resistance is a key term to estimate the bed shear stress, which is critical for many hydrodynamic modeling applications, such as flood inundation modeling, sediment transport, and river morphodynamics. For example, smaller flow resistance than the actual value will underestimate the flooding area and the sediment transport capacity, which can cause significant economic losses and environmental damages \cite{badoux2014damage,wing2022inequitable}. 

Flow resistance can be estimated by Manning’s resistance equation using Manning’s roughness coefficient, $n$. In practice, Manning’s $n$ is determined from tables, diagrams, empirical formulas, and often professional judgment, which introduces significant uncertainty in the hydrodynamic modeling results. Tremendous efforts have been made to improve the accuracy of the flow resistance parameterization by proposing unified explicit formulas for Manning's $n$ \cite{Barr1977,Churchill1973,Yen2002,Cheng2008,ChenEtAl2019}. Currently, how to properly represent and parameterize the flow resistance is still a challenge for the hydrological and hydraulic modeling community. In fact, the roughness coefficient is not only a physical parameter as we know from textbooks, but also a numerical parameter because it may also depend on mesh resolution, bathymetry data resolution, turbulence closure, and numerical schemes in different models \cite{LiuEtAl2024NCHRP}.

Inverse modeling of Manning’s $n$ using SWEs can effectively account for the spatial variation and uncertainties of flow resistance, thus providing reliable inferences. However, it is challenging to use traditional SWEs solvers for inverse modeling due to their lack of efficient differentiability for accurately capturing model sensitivity. There are many such SWEs solvers, for example SRH-2D by U.S. Bureau of Reclamation (USBR) \cite{lai2010two}, HEC-RAS 2D by U.S. Army Corp of Engineers (USACE) \cite{brunner1995hec}, FLO2D \cite{o2011flo}, RiverFlow2D \cite{hydronia2016riverflow2d}, MIKE 21 \cite{warren1992mike}, TUFLOW \cite{huxley2016tuflow}, LISFLOOD \cite{van2010lisflood}, PAWS \cite{Shen2010}, and GPU-supported TRITON \cite{TRITON2022}. As a remedy, previous studies have used the stochastic inversion method to avoid evaluating model sensitivity by constructing a search space \cite{butler2015definition,siripatana2018ensemble} or computationally inexpensive surrogate models \cite{siripatana2020bayesian,meert2018surrogate}. However, these methods have limitations due to the curse of dimensionality. It is difficult to use these methods when the number of parameters is large.  This work focuses on deterministic inverse modeling using gradient descent, for which model differentiability is essential. 

There are several ways to calculate gradients for inverse modeling. One simple way is to use the finite difference method (FDM) by perturbing parameters. However, its accuracy depends on the perturbation magnitude. A more complex approach is to use adjoint equations to derive the gradient of the optimization functions for SWEs \cite{wei2024effects}. These adjoint equations are problem specific and need to be solved together with the original governing equations. Another way is to use deep-learning-based surrogate models such as the one proposed in \citeA{song2023surrogate}. Surrogate models based on deep neural network (NN) can leverage automatic differentiation (AD) supported in machine learning platforms, such as PyTorch, TensorFlow, and JAX in Python, or SciML in Julia. This allows for efficient gradient tracking and enables gradient-based optimization for both training and inversion \cite{guo2024reduced,ohara2024physics,LiuEtAl2024,cao2024laplace}. Some surrogate models incorporate physics-constrains into the training process to enhance accuracy and generalization, such as Physics-Informed Neural Networks (PINN; \cite{raissi2019physics}). However, they still need to be pre-trained before inversion and their performance may degrade for out-of-sample scenarios.

Recently, differentiable modeling (DM) has emerged as a new approach for gradient-descent-based inversion, which can directly leverage AD in physics-based models and eliminate the need for surrogate. These models use AD to track the gradients of fundamental operations throughout the entire computational pipeline and accumulate them using the chain rule. In addition, their differentiability enables the seamless integration with NNs, where NNs can be used to learn model parameters or replace physical modules for poorly represented processes \cite{Shen2023}. DMs have demonstrated state-of-the-art performance, full interpretability, and strong extrapolation capability in out-of-sample scenarios within hydrology and geosciences \cite{jiang2020improving,Shen2023,FengEtAl2023,AboelyazeedEtAl2023,rahmani2023identifying,song2024improving,song2024high,wang2024distributed,zhong2024development}. 

In contrast, the development of DM in hydrodynamics and hydraulics is still limited. The main reason is the prevalent hydrologic and land surface models are conceptual and governed by ordinary differential equations (ODEs), while the governing equations in hydrodynamics and hydraulics are more complex partial differential equations (PDEs), such as the SWEs. Following the concept of the universal differential equation (UDE) proposed in \citeA{rackauckas2020universal}, and more importantly enabled by the differentiable programming, we combine NNs with a differentiable two-dimensional (2D) SWEs solver and propose the Universal Shallow Water Equations (USWEs). In this work, the idea of USWEs is implemented in Julia, a differentiable programming language, within the software package \Hydrograd developed by the authors for computational hydrodynamics. It supports end-to-end differentiability and scientific machine learning.

In this work, we used flow resistance as an example to demonstrate the capabilities of the differentiable USWEs solver. The flow resistance term and parameters such as the friction coefficient $f$ and Manning's roughness coefficient $n$ are difficult to be determined because of the complex flow dynamics and the large number of influencing factors. There might also be factors we simply do not know, cannot measure, or cannot represent in the models. These attributes make the flow resistance problem well suited to be solved with the USWEs approach. In USWEs, the roughness coefficient or the flow resistance term can be replaced by a NN, whose training is indirectly driven by the flow simulation and observational data. Essentially, the NN is used to capture the complex functional relationship between the flow resistance and its influencing variables.  

The critical importance of the flow resistance term and the parameters has been well recognized in the river hydraulics and hydrodynamic modeling community. From the 18th century starting from the work of Chezy \cite{Chezy1775}, to the 19th century with the work of Manning \cite{Manning1891}, to the 20th century with the work of Nikuradse \cite{Nikuradse1933}, and the 21st century with the work of Yen \cite{Yen2002} and many others, researchers have proposed many empirical formulas to estimate the flow resistance \cite{ChenEtAl2019}. In general, the flow resistance is a function of the flow velocity, water depth, bed roughness, vegetation, sediment transport and other factors. In practice, the flow resistance parameters may be treated as a constant, a function of flow depth, or a function of multiple variables such as flow velocity, water depth, and bed roughness. This work will show application examples using all three treatments. One example will demonstrate the flow field sensitivity to the distribution of constant $n$ values and the other two demonstrate the capability of the USWEs approach to learn the complex functional forms of the flow resistance. The examples will show that the USWEs approach can indeed learn and distill the flow resistance physics and knowledge from the results of more than 200 years of research and practice. With the increasingly advanced sensing technology and the availability of large datasets, this opens a new opportunity to advance the state-of-the-art in flow resistance physics discovery and modeling. 

The rest of the paper is organized as follows: Section \ref{section:framework} provides an overview of the differentiable modeling framework \Hydrograd, including its core structure, mathematical formulation, and implementation details. Section \ref{section:applications} demonstrates the applications in sensitivity analysis, parameter inversion, and flow resistance physics discovery. Section \ref{section:discussion} discusses several current issues related AD, differentiable programming, and future research needs. Section \ref{section:discussion} concludes the paper. The focus of this paper is not on the computer code development, but on the scientific and mathematical aspects of the USWEs approach and its applications. Thus, the details about the code implementation can be found in the Supplementary Information and online at \url{https://github.com/psu-efd/Hydrograd.jl}.

\section{Differentiable Modeling for the Universal SWEs} \label{section:framework}

\subsection{Structure of \Hydrograd}
\begin{figure}[htp]
    \centering
    \includegraphics[width=1\linewidth]{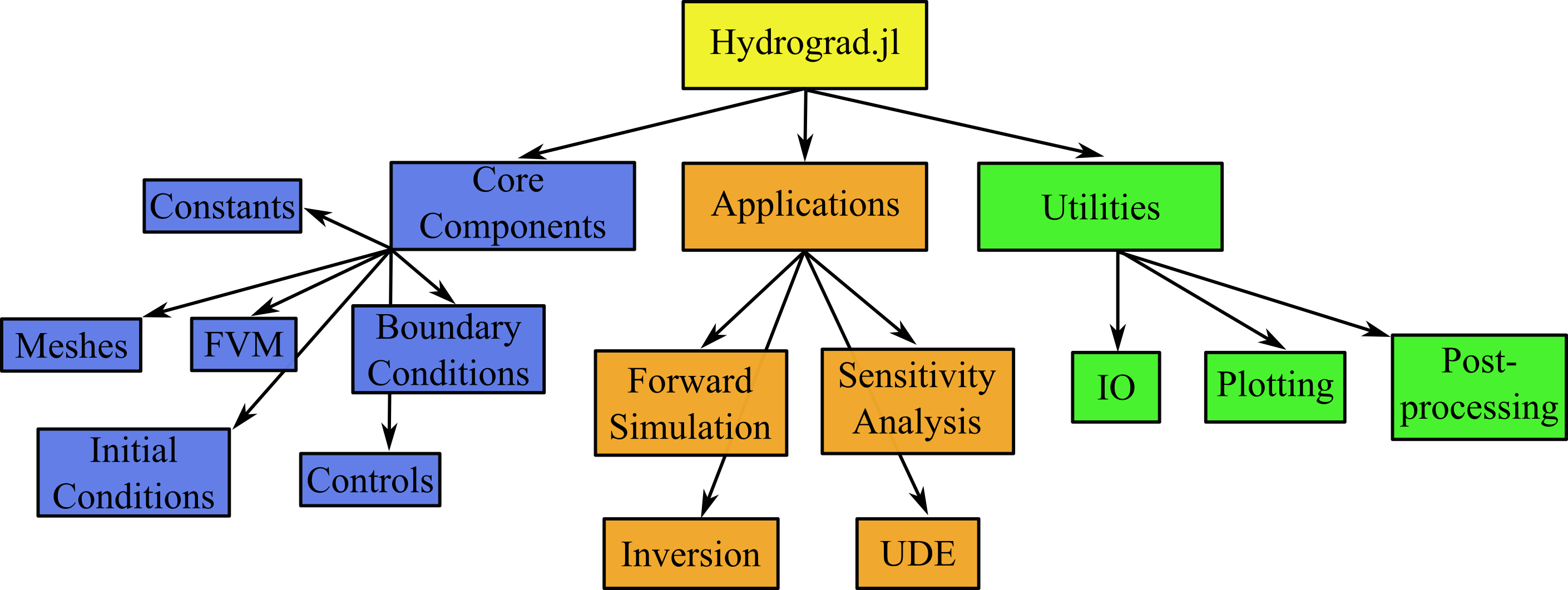}
    \caption{The code structure of \Hydrograd.}
    \label{fig:Hydrograd_structure}
\end{figure}

\Hydrograd is a modular Julia package designed for forward simulations, sensitivity analysis, parameter inversion, and scientific machine learning. The package is organized in three main sections: Core Components, Applications, and Utilities, each contributing to different aspects of the package's functionality. Figure~\ref{fig:Hydrograd_structure} shows the code structure of \Hydrograd. The modular structure ensures flexibility, reusability, and extensibility for diverse computational models and workflows. Currently, \Hydrograd only supports the implementation of 2D SWEs. Due to the modular structure, other types of PDEs in hydrodynamics and morphodynamics can also be implemented in the future without breaking the existing code. \Hydrograd can utilize the Scientific Machine Learning (SciML) ecosystem in Julia, which provides a suite of tools for scientific machine learning, including differential equation solvers, AD, adjoint-based backward functions, neural networks, and optimization \cite{rackauckas2017differentialequations}. More details about the code structure and implementation can be found in the Supplementary Information.

\subsection{Mathematical Formulation for the Universal SWEs Solver}

\subsubsection{Governing Equations}
\begin{figure}[htp]
    \centering
    \includegraphics[width=1\linewidth]{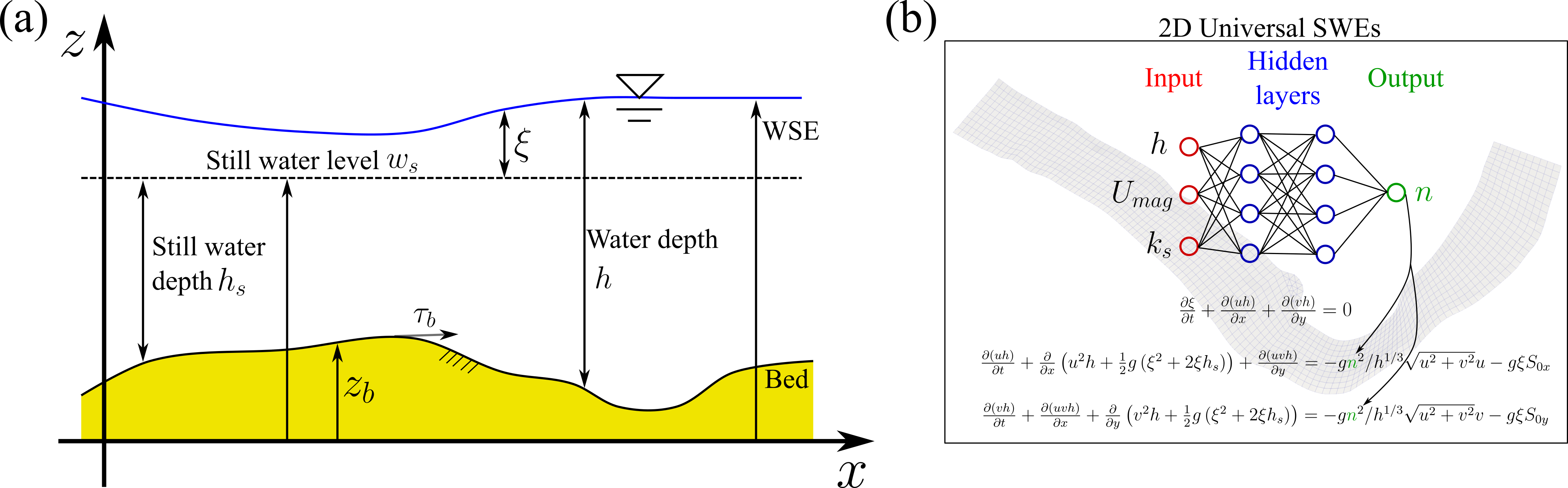}
    \caption{The schematic diagram of the shallow flow and definitions: (a) the definitions of the variables; (b) the scheme of the universal shallow water equations (USWEs).}
    \label{fig:swe_scheme_UDE}
\end{figure}

The depth-averaged SWEs are a set of partial differential equations that describe the flow in a channel with a free surface. A schematic diagram of the shallow flow and definitions are shown in Figure \ref{fig:swe_scheme_UDE}(a). The 2D SWEs solved by \Hydrograd are: 
\begin{equation}\label{eqn:cty}
    \frac{\partial \xi}{\partial t} + \frac{\partial (uh)}{\partial x} + \frac{\partial (vh)}{\partial y} = 0
\end{equation}
\begin{equation}\label{eqn:momentum_x}
    \frac{\partial (uh)}{\partial t} + \frac{\partial}{\partial x} \left( u^2 h + \frac{1}{2} g \left( \xi^2 + 2 \xi h_s \right) \right) + \frac{\partial (uvh)}{\partial y} =  - \frac{\tau_{bx}}{\rho} - g \xi S_{0x} 
\end{equation}
\begin{equation}\label{eqn:momentum_y}
    \frac{\partial (vh)}{\partial t} + \frac{\partial (uvh)}{\partial x} + \frac{\partial}{\partial y} \left( v^2 h + \frac{1}{2} g \left( \xi^2 + 2 \xi h_s \right) \right) =  - \frac{\tau_{by}}{\rho} - g \xi S_{0y} 
\end{equation}
where $\xi$ is the free surface elevation above the still water level $h_s$, $h$ is the total water depth (=$h_s$ + $\xi$),  $u$ and $v$ are the depth-averaged flow velocities in $x$ and $y$ directions, respectively. $z_b$ is the bed elevation and thus the water surface elevation ($WSE$) is $h+z_b$; $\tau_{bx}$ and $\tau_{by}$ are the bed shear stresses in $x$ and $y$ directions, respectively; $g$ is the gravitational acceleration; $\rho$ is the water density. For simplicity, other terms such as the viscous and turbulence stresses are neglected in the momentum equations. They are not critical for the purpose of this paper and can be easily added in the future. The bed shear stress terms, $\tau_{b,x}$ and $\tau_{b,y}$, and how the USWEs approximate them will be described in more details in the following sections. 

The formulations of the 2D SWEs in Equations \ref{eqn:cty}, \ref{eqn:momentum_x}, and \ref{eqn:momentum_y} are the same as those in \citeA{LiuEtAl2008}, which follow the same source term split method proposed in \citeA{RogersEtAl2001} and \citeA{RogersEtAl2003}. For example, the original momentum source term related to pressure gradient in the $x$-direction has the form of $gh\frac{\partial \xi}{\partial x}$. However, for cases with uneven bathymetry, the original form of the source term does not produce balanced momentum when the inter-cell inviscid fluxes are solved with Roe's Riemann solver, which is used in \Hydrograd. Without proper treatment, artificial momentum fluxes will be generated and the results are not physically correct. The source term split method reformulates the pressure gradient term in $x$-direction (similarly in $y$-direction) as: 
\begin{equation}\label{eqn:source_term_split}
    gh\frac{\partial \xi}{\partial x} = \frac{1}{2} g \frac{\partial \left(\xi^2 + 2 \xi h_s \right)}{\partial x} - gh\frac{\partial z_b}{\partial x}
\end{equation}
which still maintains the hyperbolic nature of SWEs and works well with Roe's Riemann solver to produce balanced momentum fluxes. The core of the idea is that the pressure gradient term is split into two parts: the first part is the pressure gradient term that is related to the free surface deviation from a reference surface, and the second part is the pressure gradient term that is related to the bed slope. When the flow is at rest, the free surface deviation $\xi$ and velocity are zero. The fluxes related to the two terms on the right-hand side of Equation \ref{eqn:source_term_split} are in perfect balance. 

\subsubsection{Flow Resistance}\label{section:flow_resistance}

The flow resistance ($\tau_{b,x}$ and $\tau_{b,y}$) is a key term in SWEs, which often are the only parameters in the calibration of hydrodynamics models. Flow resistance is a complex function of the flow velocity, water depth, bed roughness, vegetation, sediment transport and other factors \cite{LiuEtAl2024NCHRP}. Open channel flow resistance has been semi-empirically modeled from the 18th century \cite{Chezy1775,Manning1891}. The modern flow resistance formulas have their roots in pipe flows from the seminal work of Nikuradse \cite{Nikuradse1933,Nikuradse1950}, Colebrook and White \cite{ColebrookWhite1937}, Moody \cite{Moody1944}, and the work in river hydraulics by Rouse \cite{Rouse1965} and many others \cite{Yen2002}.

Flow resistance is often parameterized by the the dimensionless Darcy-Weisbach friction coefficient $f$ and the Manning's roughness coefficient $n$. The bed stresses, $\tau_{bx}$ and $\tau_{by}$, can be estimated by the Manning's resistance equation as:
\begin{equation}\label{eqn:Manning_resistance}
(\tau_{bx}, \tau_{by}) = \frac{\rho}{8} f \sqrt{u^2+v^2}(u,v)
\end{equation}
where $f = 8g n^2/h^{1/3}$. The two parameters $f$ and $n$ are related to each other, i.e., one can be computed from the other. 

In this work, two categories of flow resistance formulas are considered. The first category is based on the fact that the Manning's $n$ is a function of flow depth. Many hydrodynamics models such as SRH-2D and HEC-RAS have the option to specify the functional relationship between Manning's $n$ and flow depth. In general, without the effect of vegetation, Manning's $n$ has a higher value for shallow flows and a lower value for deep flows. Under the condition of very shallow flows or sheet flows such as overland runoff, the form-drag on roughness elements plays the dominant role in the flow resistance. When the flow is deep, the roughness height of the bed material is less significant and the Manning's $n$ asymptotically approaches a constant value. In the literature, there are empirical formulas to describe such functional dependency of Manning's $n$ on flow depth, such as the Strickler formula \cite{hager2015albert}, Manning-Strickler formula \cite{yen1992dimensionally}, Limerinos' equation \cite{limerinos1970determination}, Bathurst's equation \cite{bathurst1985flow}, and Cowen's Equation \cite{marcus1992evaluation}. In this work, as an attempt to generalize these functional relationships, a simple Sigmoid function was adopted, which has the form of 
\begin{equation}\label{eqn:sigmoid_n_h}
    n(h) = n_{\text{lower}} + \frac{ n_{\text{upper}} - n_{\text{lower}} }{1 + e^{-k(h-h_{\text{mid}})}}   
\end{equation}
where $n_{\text{lower}}$ and $n_{\text{upper}}$ are the lower and upper bounds of the Manning's $n$ values for deep and shallow flow depth, respectively, and $k$ and $h_{\text{mid}}$ are the parameters that control the shape and the transition point of the function. When the water depth is at $h_{\text{mid}}$, the Manning's $n$ value is simply the average of $n_{\text{lower}}$ and $n_{\text{upper}}$. For some application examples in this paper, in the forward simulation to generate the ground truth for training, this Sigmoid function was used to compute the Manning's $n$ value for each cell of the mesh given the local water depth. As a result, the Manning's $n$ values are not constant and they are spatially varying in the channel. 

The second category is based on the Darcy-Weisbach friction factor $f$ where not only the flow depth, but also the velocity and the relative roughness height of the bed material are considered. In the authors' opinion, $f$ is more generalizable than the Manning's $n$ because $f$ is dimensionless while the latter is not. However, the use of Manning's $n$ is prevailing currently. In general, $f$ is a function of the Reynolds number $Re$ and the relative roughness height $h/k_s$ where $k_s$ is the roughness height. \citeA{Cheng2008} proposed a unified explicit formula for $f$ as:
\begin{equation}\label{eqn:Cheng_f_Re_hks}
    \frac{1}{f} = \left( \frac{Re}{24} \right)^{\alpha} \left(1.8 \log \frac{Re}{2.1} \right)^{2(1-\alpha)\beta}  \left( 2 \log \frac{11.8h}{k_s} \right)^{2(1-\alpha)(1-\beta)}
\end{equation}
where $Re = U_{mag} h/\nu$ is the Reynolds number, $U_{mag}$ (=$\sqrt{u^2+v^2}$) is the velocity magnitude, and $\nu$ is the kinematic viscosity of the fluid. The parameters $\alpha$ and $\beta$ are defined as
\begin{equation}
    \alpha = \frac{1}{1 + \left(Re/850 \right)^9}
\end{equation}
and
\begin{equation}
    \beta = \frac{1}{1 + \left(Re/(160h/k_s) \right)^2}.
\end{equation} 
Cheng's formula is applicable across the whole range of $Re$ and $h/k_s$. Thus, it is adopted in this work for some of the application examples. 

In \Hydrograd, the Manning's $n$ in a forward simulation can be a constant (though allowing variations in different roughness zones), a function of flow depth using Equation \ref{eqn:sigmoid_n_h}, or a function of flow depth, velocity, and roughness height using Equation \ref{eqn:Cheng_f_Re_hks}. As will be shown in the application example section, the forward simulations with these flow resistance options can be performed to generate the training data for parameter inversion and USWEs-based discovery of flow resistance physics.

\subsection{Sensitivity Analysis and Parameter Inversion}
The differentiable USWEs solver can be written as:
\begin{equation}\label{eqn:differentiable_hydrodynamics}
    \mathbf{Q}_{\text{pred}}(t) = \mathbf{G(t,\mathbf{\theta})}  
\end{equation}
where $\mathbf{G}$ represent the forward model, $\mathbf{Q}_{\text{pred}}$ is the model prediction of the solution vector ($[\xi, hu, hv]$) defined in each cell of the mesh and $\mathbf{\theta}$ is the parameter vector, which encapsulates all model inputs and parameters such as Manning's $n$, bed elevation $z_b$, initial and boundary conditions. In addition, when a NN is in the USWEs, the weights of the NN is also part of the parameter vector $\mathbf{\theta}$. The sensitivity of the model prediction vector $\mathbf{Q}_{\text{pred}}$ with respect to the parameter vector $\mathbf{\theta}$ is the Jacobian matrix ${\partial \mathbf{G}}/{\partial \mathbf{\theta}}$. 

Parameter inversion often involves the optimization of the parameter vector $\mathbf{\theta}$ to minimize a prediction loss function measuring the difference between the model prediction $\mathbf{Q}_{\text{pred}}$ and the ground truth data $\mathbf{Q}_{\text{obs}}$. For example, the $L_2$ norm of the difference, $L_{\text{pred}} = \left\| \mathbf{Q}_{\text{pred}} - \mathbf{Q}_{\text{obs}} \right\|^2$, is often used. For the case of SWEs, the prediction loss can further be decomposed into the loss due to the water surface elevation mismatch $L_{\text{WSE}}$ and the loss due to the flow velocity mismatch $L_{\text{uv}}$. For some parameters, regularization terms, noted as $L_{\text{reg}}$, are needed as additional constraints to the loss function to prevent overfitting. This is a common practice in parameter inversion due to the ill-posed nature of the inverse problem, especially in high-dimensional parameter spaces. \Hydrograd supports the zeroth-order Tikhonov regularization for the parameter range and the first-order Tikhonov regularization for the parameter gradient \cite{Aster2013}. Regularization is necessary for the success of inversion as demonstrated in \citeA{LiuEtAl2024} for the case of bathymetry inversion. The total loss $L_{\text{total}}$ is the sum of the prediction loss $L_{\text{pred}}$ and the regularization term $L_{\text{reg}}$ as:
\begin{equation}\label{eqn:total_loss}
    L_{\text{total}} = L_{\text{pred}} + \lambda L_{\text{reg}}
\end{equation}
where $\lambda$ is the regularization parameter which weights the contributions of the prediction loss $L_{\text{pred}}$ and the regularization term $L_{\text{reg}}$. 

Due to the differentiability of the USWEs solver, the gradient of the total loss $L_{\text{total}}$ with respect to the parameter vector $\mathbf{\theta}$ can be automatically computed, which can then be ingested by gradient-based optimization algorithms. \Hydrograd supports the gradient-based optimizers in ``Optimization.jl'' \cite{optimization2023}. In the examples shown in this paper, the ADAM optimizer is used for the inversion. Note that the inversion is often performed for a subset of the full model parameter vector $\mathbf{\theta}$. Currently, \Hydrograd supports the sensitivity analysis and inversion of Manning's $n$ values, bathymetry $z_b$, boundary conditions such as the upstream inflow discharge, and the weights of NN. For parameter inversion, proper normalization of the loss functions is critical for the optimization to converge. In this work, the min-max normalization is used for the loss functions, i.e., the solution variables are normalized to the range of $[0, 1]$.  

\subsection{Scientific Machine Learning Integration}
A key innovation within scientific machine learning is the UDE idea, which embeds machine learning components (e.g., NN) directly into differential equations \cite{rackauckas2020universal}. This concept is adopted in this work to create USWEs which integrate the physics-based SWEs with deep-learning. What makes this integration possible is the differentiability of both the SWEs solver and NNs, which allows the gradients to propagate throughout the computational pipeline. USWEs allow for data-driven discovery of unknown or partially known dynamics such as flow resistance laws by learning from data while retaining the structure and constraints of the underlying flow physics in the form of 2D SWEs. 

As a proof-of-concept, this paper shows how to use USWEs to learn the flow resistance laws described in Section~\ref{section:flow_resistance}, i.e., the functional relationship between Manning's $n$ and water depth $h$, flow velocity, and roughness height $k_s$. Figure~\ref{fig:swe_scheme_UDE}(b) shows the idea of USWEs where a NN is used to learn the functional relationships. The figure shows the NN has three inputs, i.e., water depth $h$, velocity magnitude $U_{mag}$, and roughness height $k_s$. The single output of the NN is the Manning's $n$. During the simulation at each time step, the water depth, flow velocity, and roughness height of each cell is fed into the NN to obtain the corresponding Manning's $n$. It is noted that for the case when the Manning's $n$ is a function of only $h$, there is only one input for the NN. 

In addition to the flow resistance laws, the idea of USWEs can be extended to learn any unknown flow dynamics or parameters that can be plugged into the SWEs, i.e., any part of the governing equations can be replaced by a NN. For example, instead of learning the Manning's $n$, one can even replace the whole flow resistance terms $\tau_{bx}$ and $\tau_{by}$ with a neural network, which can take flow velocity, water depth, bed roughness, vegetation, and other factors as its inputs and outputs the bed shear stresses directly. 


\subsubsection{Numerical Schemes}
The governing equations of the USWEs are solved for the conservative variables $\mathbf{Q}$ using the method of lines \cite{schiesser1991method}. The SWEs are spatially discretized on unstructured meshes using the finite volume method (FVM) to convert the PDEs into a set of ODEs and then the ODEs are solved using a time-marching scheme. The numerical schemes for spatial discretization generally follows those in \citeA{LiuEtAl2008}. To reduce the length of the paper, the details of the numerical schemes are not presented here. In essence, the spatial discretization computes the inter-cell inviscid fluxes using the Roe's Riemann solver. Along a face between two cells, the solution variables ($\mathbf{Q}$) are interpolated to both sides of the face and the fluxes are the solution of the Riemann problem. The Roe's Riemann solver, a popular high-order approximate solver, is implemented in \Hydrograd. After spatial discretization, the resulting ODEs can be written in the form of an ODE system:
\begin{equation}\label{eqn:ODEs}
    \frac{d\mathbf{Q}}{dt} = \mathbf{f}(\mathbf{Q}, \mathbf{\theta})
\end{equation}
where $\mathbf{Q}$ is the solution vector defined in each cell of a mesh and $\mathbf{\theta}$ is the parameter vector. The ODE system above is usually very stiff due to the fact that it is resulted from the spatial discretization of a hyperbolic PDE system, the time-marching scheme must be suitable for such a system. SciML's ``DifferentialEquations.jl'' package is used to solve the ODEs, which provides a large selection of solvers for stiff ODEs. For all the cases presented in this work, the Tsitouras 5/4 Runge-Kutta method \cite{Tsitouras2011} was used and no instability was observed.  

For sensitivity analysis, inversion, and the training of USWEs, a significant portion of the computing time is spent on obtaining the gradient and Jacobian of the model prediction vector $\mathbf{Q}_{\text{pred}}$ with respect to the parameter vector $\mathbf{\theta}$. The chain rule of differentiation is a fundamental concept in calculus that allows us to find the derivative of a composite function such as the loss function defined in Equation~\ref{eqn:total_loss} by breaking it down into simpler functions whose derivatives can be analytically computed. There are multiple ways to automatically compute the gradient and Jacobian. These methods can be broadly categorized into forward-mode AD, which propagates derivatives forward through the computation graph and is efficient for functions with few inputs and many outputs, and reverse-mode AD, which propagates derivatives backward and is efficient for functions with many inputs and few outputs. Source-code transformation and operator overloading are two main implementation approaches for AD, each with different tradeoffs in terms of memory usage and computational efficiency. It is beyond the scope of this paper to discuss the details of each method. Interested readers can refer to \citeA{sapienza2024differentiable} for an in-depth review. A short summary is provided in the Supporting Information of this paper. In \Hydrograd, both the forward-mode and reverse-mode AD are implemented and the user can choose the one that is most efficient for the problem at hand.

The differentiable USWEs solver in \Hydrograd is not only for academic research but also for real-world applications. Thus, the FVM discretization is on a 2D unstructured mesh. Real-world applications often have complex geometries, such as river confluences, meandering channels, and complex cross-sections. The unstructured mesh allows for the flexibility to model these complex geometries. In addition, \Hydrograd directly supports the case setup of the USBR SRH-2D model, including its mesh files and other input files. Thus, \Hydrograd can directly use the pre- and post-processing tools of SRH-2D. In the future, the support for using the input files of other popular hydrodynamic models, such as HEC-RAS which similarly uses unstructured meshes, can be added through mesh conversion tools such as those in the open-source Python package pyHMT2D \cite{pyHMT2D}. 

\section{Application Examples} \label{section:applications}

This section presents the validation and applications of the differentiable USWE solver in forward simulation, sensitivity analysis, parameter inversion, and physics discovery for flow resistance. There are two example cases used. One is a classical 1D validation case for SWEs solvers and the other is a real-world case of the Savannah River in the U.S. 

\subsection{Forward Modeling Validation}\label{subsection:forward_modeling_validation}
The SWEs solver is firstly validated with a simple classical 1D case of steady flow over a bump \cite{goutal1997proceedings}. The river channel is 25-m long with a bump defined as follows:
\begin{equation}
    z_b(x) =
\begin{cases} 
0.2 - 0.05(x - 10)^2 & \text{if } 8 < x < 12, \\
0 & \text{otherwise}.
\end{cases}
\end{equation}
where $x$ is the distance along the channel. The bump is 0.2 m high and 4 m long. With an upstream inflow of 0.18 m$^2$/s and a downstream water depth of 0.33 m, the channel will have a transcritical flow with a shock. Analytical solution was given in \citeA{goutal1997proceedings} for the case of inviscid flow and ignoring the flow resistance. 

In this work, the 1D flow was simulated with a 2D mesh where the channel has a width of 1 m and only one cell in the $y$ direction. The left and right bank boundary conditions were set as frictionless walls. In the $x$ direction, the mesh has 200 cells of uniform size. To properly regulate the solution, reduce the discontinuity due to the shock, and make the case more realistic, the flow resistance term was included with a uniform Manning's $n$ of 0.02. In this way, the shock (discontinuity) is removed to ensure differentiability of the solution. The flow was simulated for 100 s to reach equilibrium. The same case was also simulated with SRH-2D for comparison. The results from \Hydrograd and SRH-2D, as well as the analytical solution, are shown in Figure~\ref{fig:oneD_channel_with_bump_UDE_result}. In general, the simulated WSE profiles are in good agreement. The profiles show the shock (hydraulic jump) over the bump. The shocks from the numerical models are less prominent than that from the analytical solution and they are also less sharp due to the flow resistance. 

\begin{figure}[htp]
    \centering
    \includegraphics[width=0.5\linewidth]{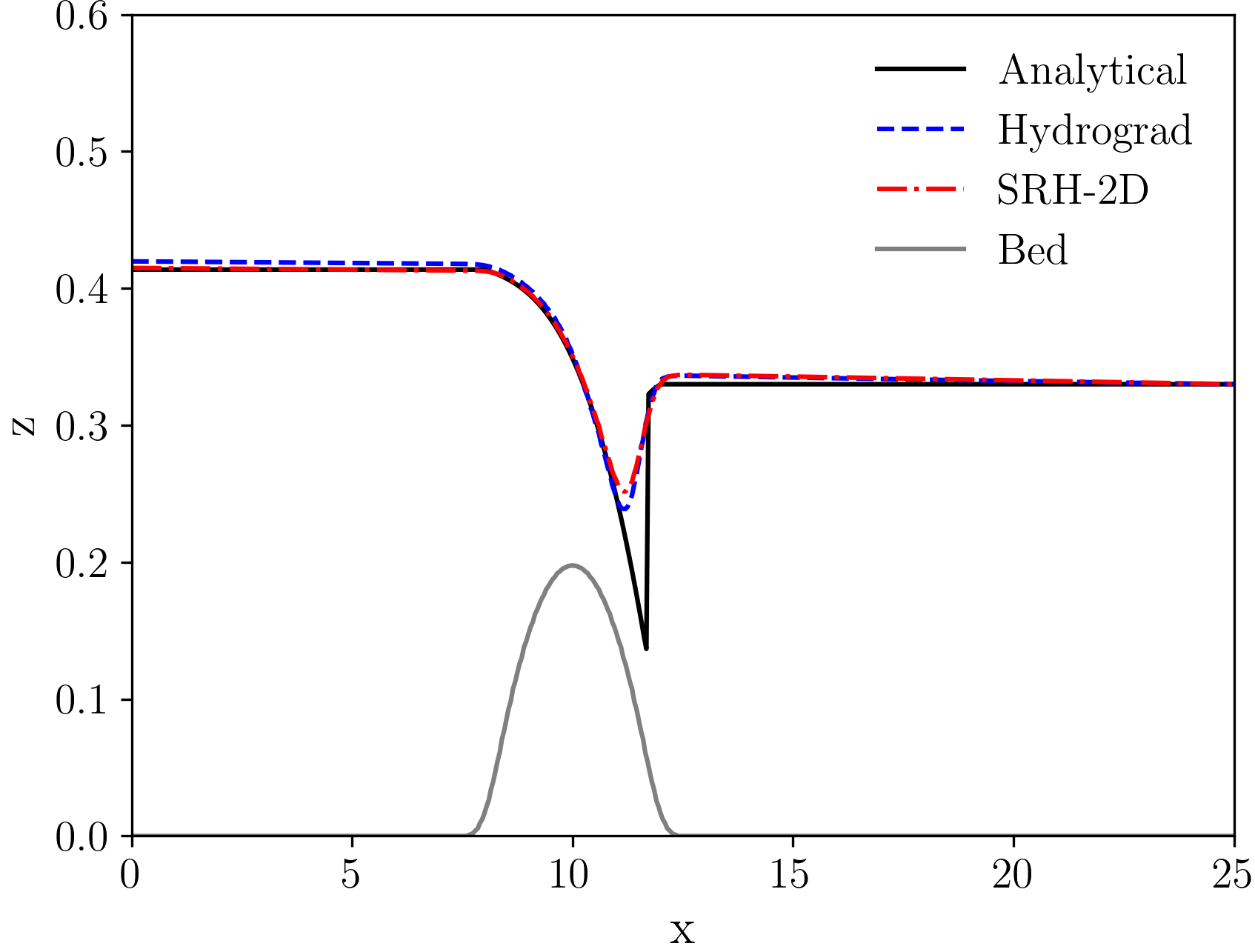}
    \caption{The comparison of the forward simulation results of the 1D channel with a bump.}
    \label{fig:forward_simulation_oneD_channel_with_bump}
\end{figure}

\subsection{Sensitivity Analysis}\label{subsection:sensitivity_analysis}

\begin{figure}[htp]
    \centering
    \includegraphics[width=1\linewidth]{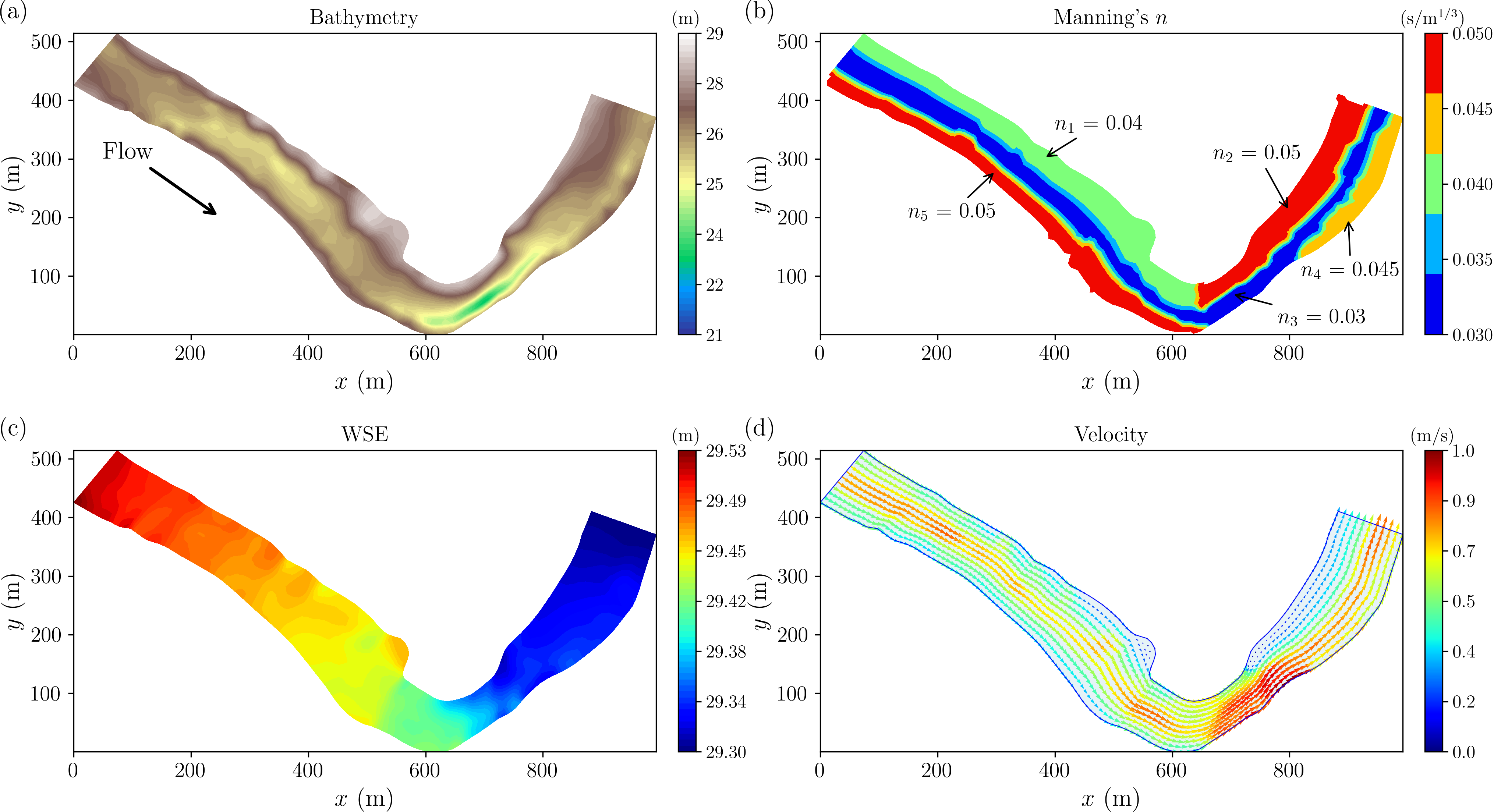}
    \caption{The forward simulation results of the Savannah River case: (a) Bathymetry, (b) Five zones of Manning's $n$, (c) Simulated WSE, (d) Simulated flow velocity.}
    \label{fig:forward_simulation_Savna}
\end{figure}

Sensitivity analysis of USWEs to flow resistence is demonstrated with the real-world case of the Savannah River near Augusta, Georgia in the U.S. The data was reported in \citeA{LeeEtAl2018} and used in \citeA{LiuEtAl2024} for bathymetry inversion. The river section is about 1.2 km long and 100 m wide. This river reach is located about 7.6 km downstream of the Savannah Bluff Lock and Dam. It features a 90$^\circ$ bend. The U.S. Army Corps of Engineers (USACE) conducted a field survey and reported the bathymetry for the river reach. The bathymetry is shown in Figure~\ref{fig:forward_simulation_Savna}(a) and the flow is from left to right. The same 2D mesh and SRH-2D model case setup used in \citeA{LiuEtAl2024} were used in this work. The mesh has 1,306 polyhedral cells and 1,430 nodes. The average cell size is about 10 m. The upstream boundary condition is a discharge of 187.4 m$^3$/s. The downstream boundary condition is a water level of 29.3 m. Since we are only interested in the sensitivity of the steady-state flow field, the flow was firstly simulated using SRH-2D to steady state. The steady state flow field from SRH-2D was then used as the initial condition for the \Hydrograd simulation, which was performed for 200 s for the flow to make small adjustment until equilibrium. 

For demonstration purpose, the whole domain is hypothetically divided into five zones with different Manning's $n$ values (see Figure~\ref{fig:forward_simulation_Savna}(b)). Zones 1 and 2 are located on the left side of the river. Zones 4 and 5 are located on the right side. Zone 3 is the middle part of the domain, which mimics the main channel. This fashion of roughness zone division is common in the field of fluvial hydraulics because the roughness characteristics of the main channel and the floodplain are often different. In the sensitivity analysis, the Manning's $n$ value for each zone is constant, i.e., it does not depend on water depth, flow velocity, or roughness height. The simulated WSE and flow velocity are shown in Figure~\ref{fig:forward_simulation_Savna}(c) and (d), respectively. 

\begin{figure}[htp]
    \centering
    \includegraphics[width=1\linewidth]{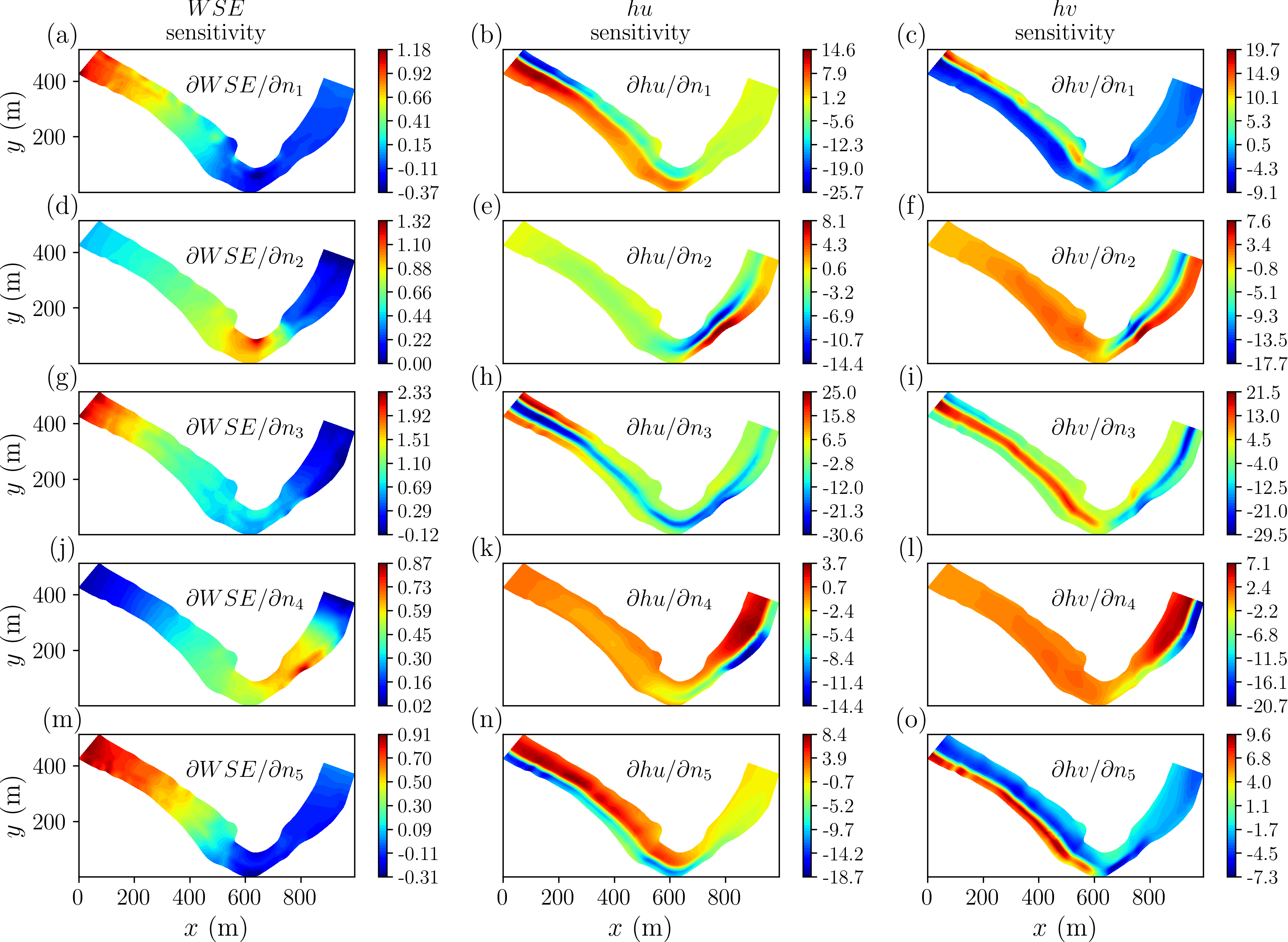}
    \caption{The sensitivity analysis of the Savannah River case. Each subplot shows the sensitivity of the flow variables ($WSE$, $hu$, $hv$) to the Manning's $n$ in different zones. Each column is for a flow variable and each row is for a roughness zone.}
    \label{fig:sensitivity_Savana}
\end{figure}

The sensitivity of the model solution $\mathbf{Q}$ with respect to the Manning's roughness coefficient $\mathbf{n}=[n_1, n_2, n_3, n_4, n_5]$, i.e., the Jacobian matrix $\mathbf{J} = \partial \mathbf{Q} / \partial \mathbf{n}$, is computed with forward-mode AD. It is suitable for this case because the number of parameters (roughness zones) is 5, which is much smaller than the length of the solution vector $\mathbf{Q}$ (3,918 = 1,306 cells $\times$ 3 variables). The sensitivity results are shown in Figure~\ref{fig:sensitivity_Savana}. Each subplot shows the sensitivity of the flow variables ($WSE$, $hu$, or $hv$) to Manning's $n$ values in different zones. The sensitivity contains both positive and negative values. The positive values indicate that the flow variable increases with Manning's $n$ value. The negative values indicate the opposite. For example, Figure~\ref{fig:sensitivity_Savana}(g) shows the WSE's sensitivity with respect to $n_3$, i.e., $\mathbf{J}[1,3] = \partial WSE / \partial n_3$. It measures how the $WSE$ changes with respect to the main channel roughness $n_3$. The majority of this sensitivity is positive, which makes physical sense because higher flow resistance slows down the flow and increase the water depth. In addition, this sensitivity is higher in the upstream part than in the downstream part. This is because the flow is subcritical and the effect of the flow resistance accumulates in the direction of from downstream to upstream, a well-known fact in open-channel hydraulics. The analysis of other sensitivities is omitted here for brevity. 

\subsection{Parameter Inversion}\label{subsection:parameter_inversion}

The Savannah River case shown above for the sensitivity analysis is also used to demonstrate parameter inversion. The constant Manning's $n$ values in the five zones are the parameters to be inverted from the training data. The training data is the $WSE$ and flow velocity simulated with the truth values of Manning's $n$ for the five zones and is shown in Figure~\ref{fig:forward_simulation_Savna}(c) and (d). For the inversion, the objective function is the total loss defined in Equation~\ref{eqn:total_loss} with $\lambda$ = 1.0. Only the zeroth-order Tikhonov regularization was used for the inversion, which penalizes the exceedance of Manning's $n$ values beyond a specified range. For this case, the range of Manning's $n$ values was set to be $[0.01, 0.06]$. In practice, the range of Manning's $n$ values can be determined by the physical characteristics of the river reach, i.e., using this as a way to infuse the prior knowledge of the Manning's $n$ values. To start the inversion, the Manning's $n$ values were initialized to be 0.03 for all zones. The ADAM optimizer with an initial learning rate of 0.001 was used to perform 300 iterations of inversion. The sensitivity (gradients and Jacobian) needed by the optimizer was computed with the forward-mode AD because of the small number of parameters (five in this case). 

\begin{figure}[htp]
    \centering
    \includegraphics[width=0.6\linewidth]{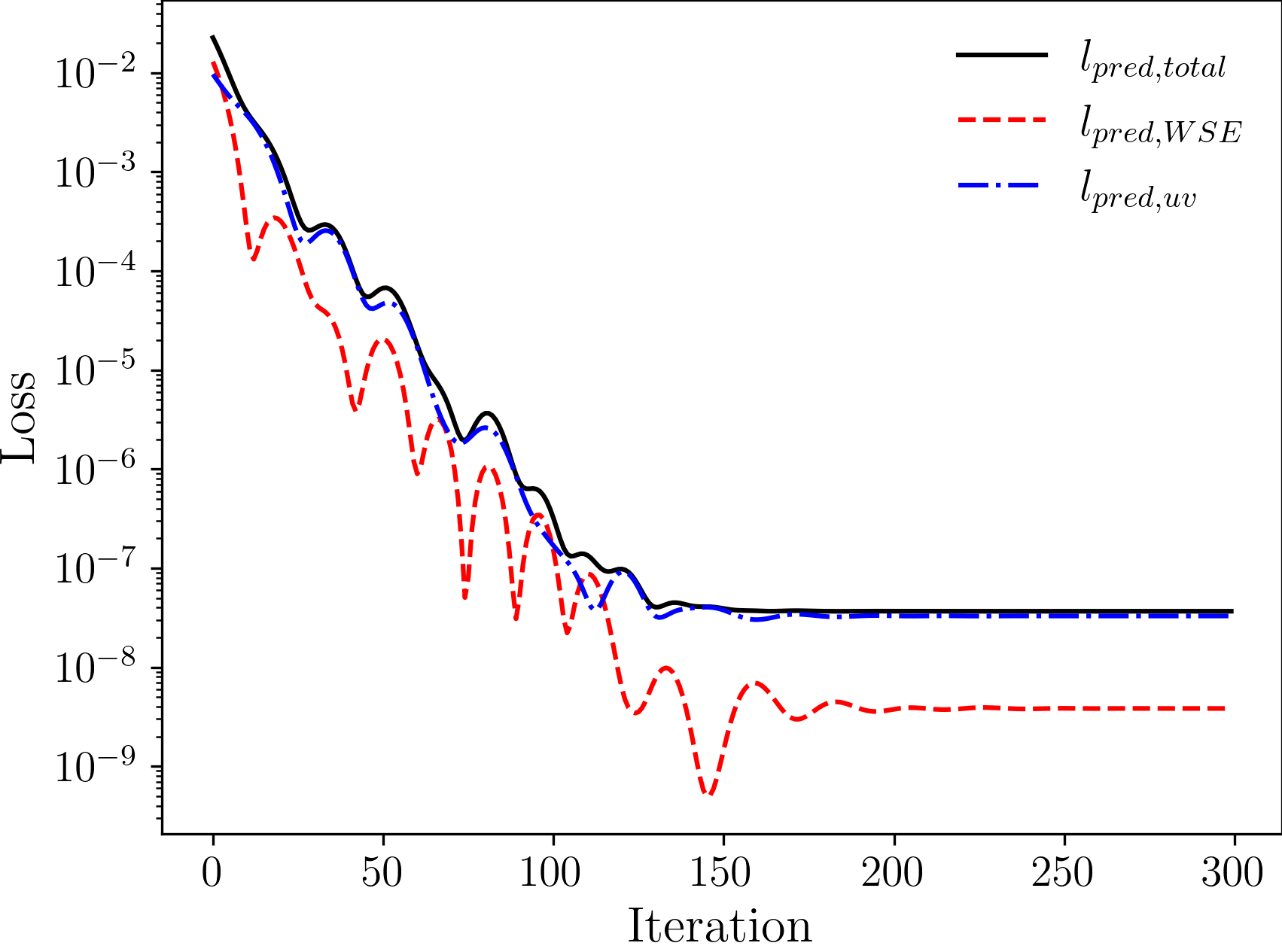}
    \caption{The histories of losses for the Manning's $n$ inversion for the Savannah River case.}
    \label{fig:inversion_losses_history}
\end{figure}

\begin{figure}[htp]
    \centering
    \includegraphics[width=1\linewidth]{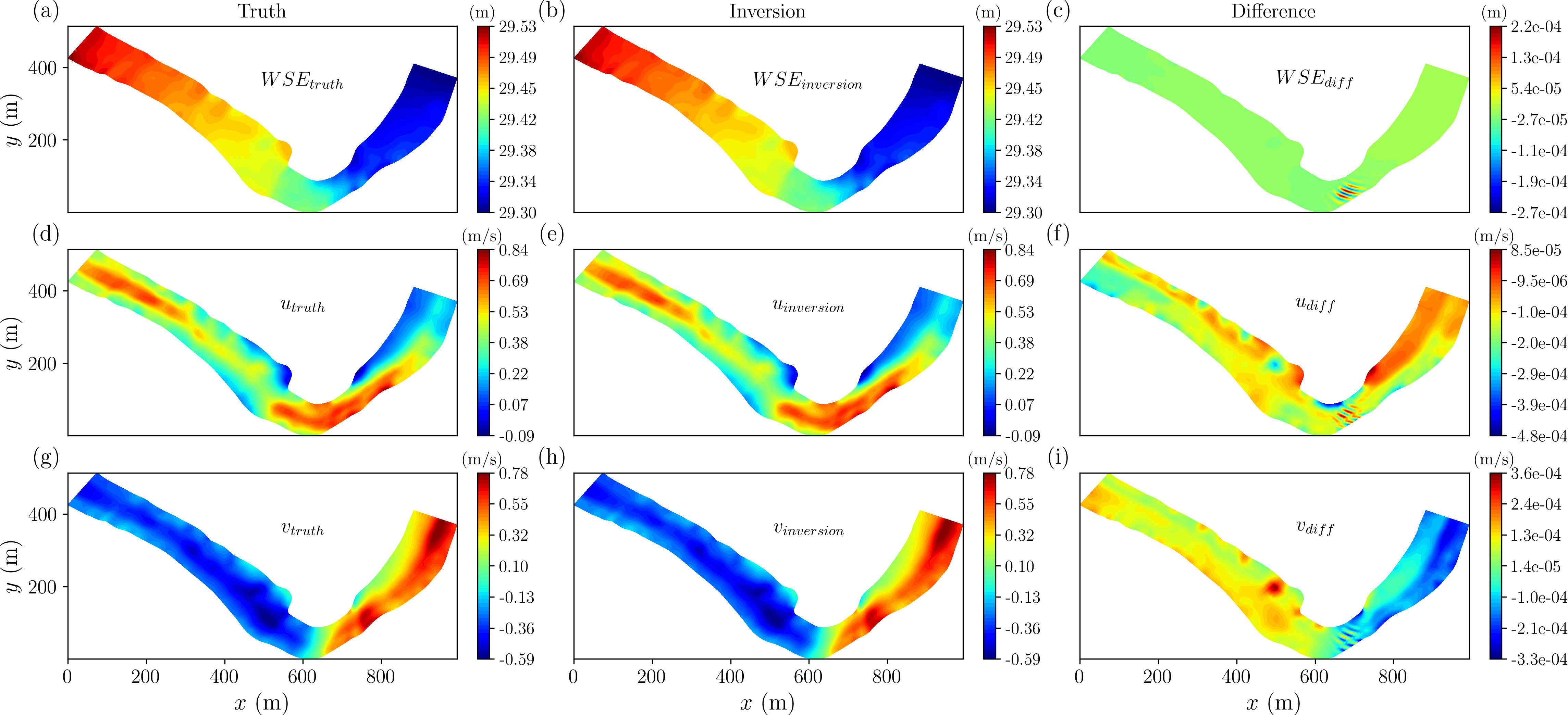}
    \caption{The comparison of the simulated flow field from the final inversion step with the observed data for the Savannah River Case. The first column shows the truth (training data) and the second column shows the simulated flow field with the final inverted Manning's $n$ values. The third column shows the difference between the simulated flow field and the truth.}
    \label{fig:inversion_Savana_flow_filed_comparison}
\end{figure}

The history of the losses is shown in Figure~\ref{fig:inversion_losses_history}. The losses were reduced by more than five orders of magnitude after about 150 iterations and then stagnated. When the inversion ended, the main contribution to the loss was from the velocity mismatch $L_{\text{uv}}$. The small final loss values indicate that the inversion was successful. The simulated flow field with the final inverted Manning's $n$ values is shown in Figure~\ref{fig:inversion_Savana_flow_filed_comparison} and compared with the truth (training data). The difference between the simulated flow field and the truth is also computed and plotted. The differences in $WSE$ and velocity components are all small and in the order of 10$^{-4}$. 

\begin{figure}[htp]
    \centering
    \includegraphics[width=1\linewidth]{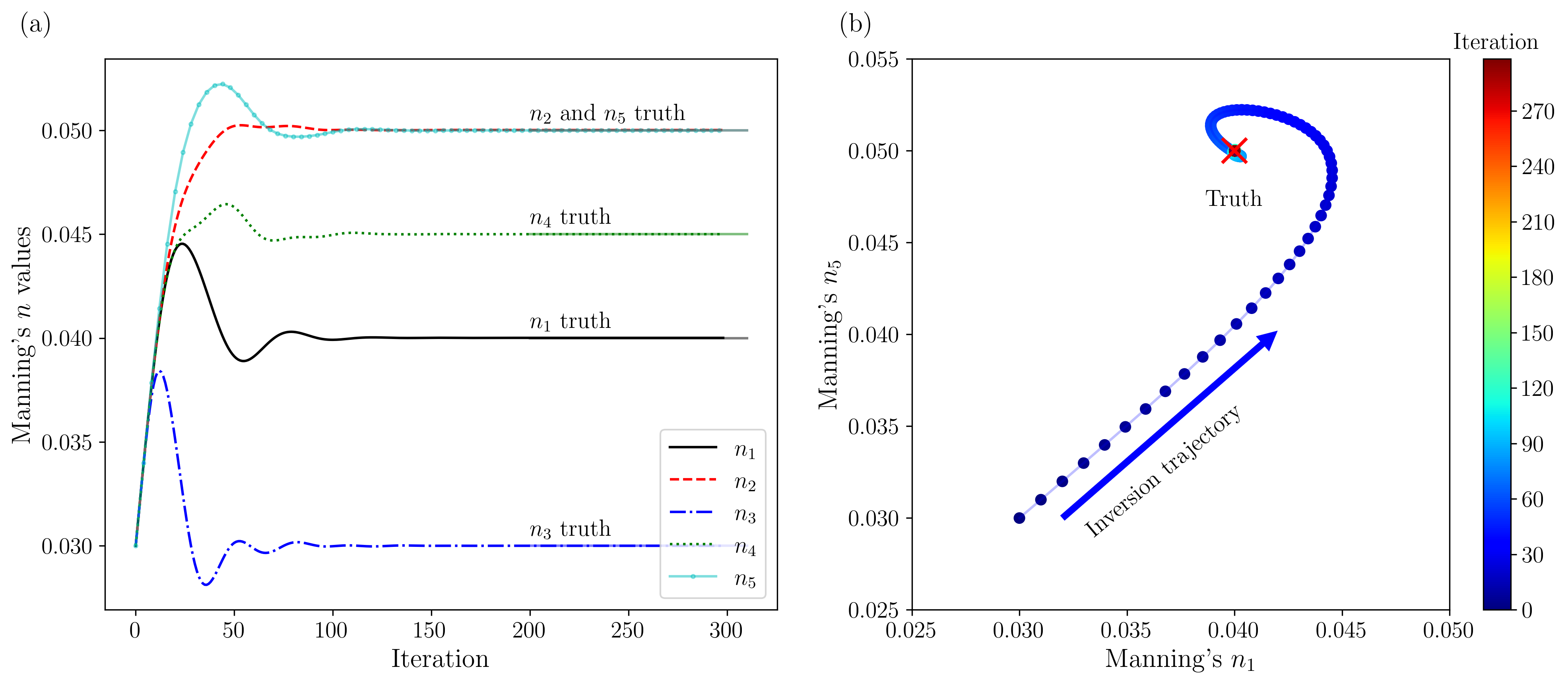}
    \caption{Inversion of Manning's $n$ values: (a) The history of the losses for the Manning's $n$ inversion, (b) The trajectory of the Manning's $n$ during the inversion process (only the pair of $n_1$ and $n_5$ is shown).}
    \label{fig:inversion_ManningN_histories_trajectories}
\end{figure}

The inversion process can be visualized by plotting the trajectories of the Manning's $n$ values. Figure~\ref{fig:inversion_ManningN_histories_trajectories}(a) shows the history of  the Manning's $n$ values during the inversion process. All the Manning's $n$ values asymptotically converged to the truth values. The trajectory of the Manning's $n$ values for $n_1$ and $n_5$ in the $n_1-n_5$ parameter space are shown in Figure~\ref{fig:inversion_ManningN_histories_trajectories}(b). The trajectory shows how the optimizer moves the pair from the initial location toward the final solution, which is very close to the truth.The smooth trajectory shows that the inversion was stable and the optimizer did not get stuck in a local minimum. This might not be true for cases where the parameter space is large, the landscape of the loss function is complex, and/or the initial guess is far away from the truth. 

\subsection{Flow Resistance Physics Discovery with USWEs}\label{subsection:physics_discovery_with_scientific_machine_learning}
In this section, two cases will be shown to demonstrate how to use USWEs and scientific machine learning to discover the two flow resistance laws in Equations~\ref{eqn:sigmoid_n_h} and \ref{eqn:Cheng_f_Re_hks}.

\subsubsection{One-Dimensional Channel with a Bump for $n(h)$}

The case of 1D channel with a bump shown in Section~\ref{subsection:forward_modeling_validation} is used to demonstrate the use of USWEs to recover the flow resistance law in Equation~\ref{eqn:sigmoid_n_h}. Unlike in Section~\ref{subsection:forward_modeling_validation}, the Manning's $n$ is now a function of water depth $h$. The training data ($WSE$, $u$, and $v$) was produced by simulating the case where the Manning's $n$ was computed from the Sigmoid function in Equation~\ref{eqn:sigmoid_n_h} with $n_{\text{lower}} = 0.03$, $n_{\text{upper}} = 0.06$, $k = 100$, and $h_{\text{mid}} = 0.3$. During the training process, the Manning's $n$ was computed with a NN which was a simple multi-layer perceptron (MLP) with two hidden layers. Each hidden layer has three neurons and the ``tanh'' activation function was used. As a proof of concept, such a small NN was sufficient for this case. For real-world applications, depending how complex the functional relationship is, a suitable NN with proper complexity can be determined through hyper-parameter tuning. During the inversion (training), the Manning's $n$ values were updated from NN and used in the SWEs solver. The objective of the training was to minimize the difference between the computed flow field and the training data, specifically $L_{\text{WSE}}$ and $L_{\text{uv}}$, by adjusting the weights in the NN. The ADAM optimizer with an initial learning rate of 0.1 was used for 100 training iterations, which is sufficient for this case. 

\begin{figure}[htp]
    \centering
    \includegraphics[width=1\linewidth]{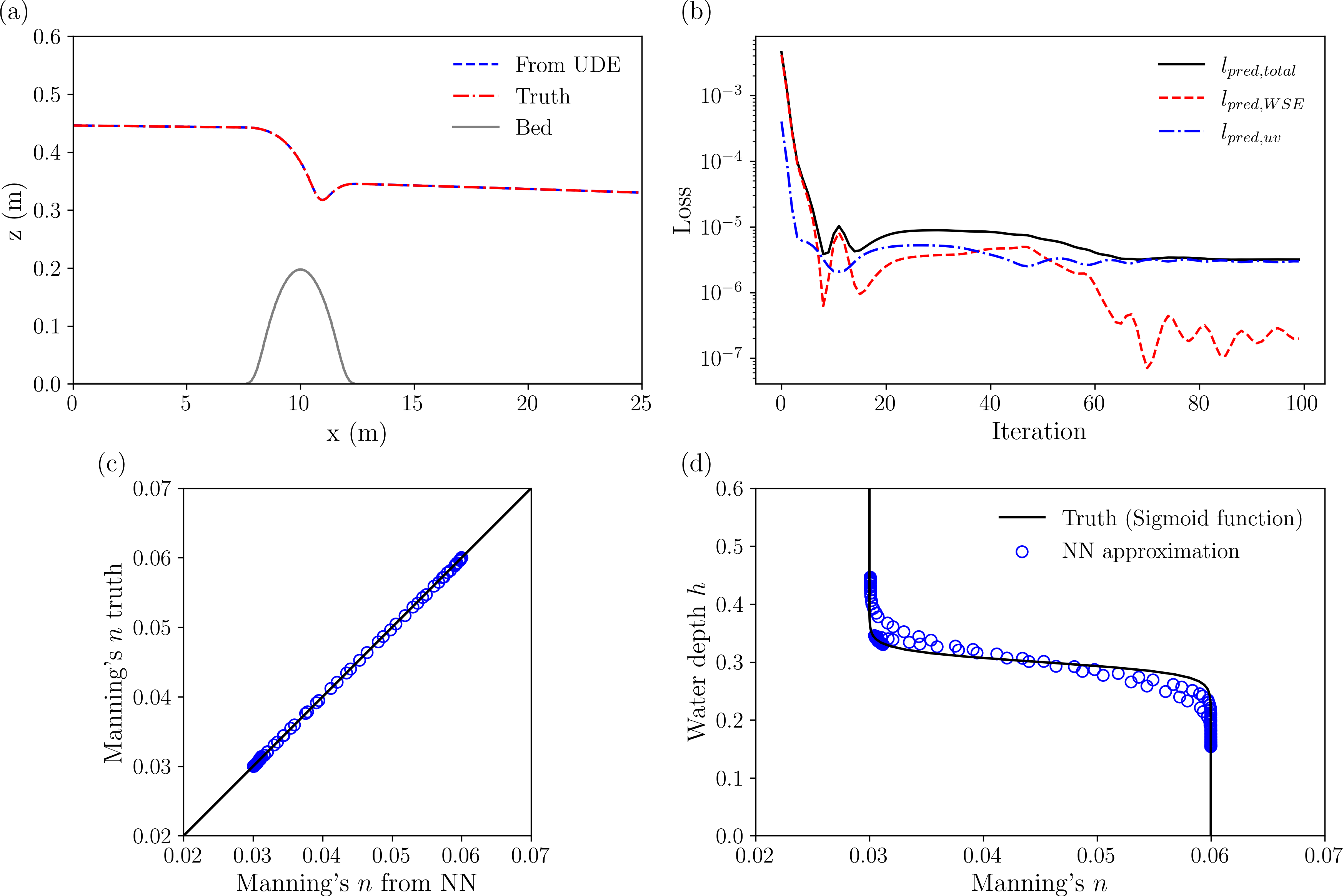}
    \caption{The result of the one-dimensional channel with a bump using USWEs: (a) Comparison of inverted WSE profile with the truth, (b) Training loss history, (c) Comparison of inverted Manning's $n$ values with the truth, and (d) Comparison of the NN prediction of Manning's $n$ with the truth of the Sigmoid function in Equation~\ref{eqn:sigmoid_n_h}.}
    \label{fig:oneD_channel_with_bump_UDE_result}
\end{figure}

The results are shown in Figure~\ref{fig:oneD_channel_with_bump_UDE_result}. The final $WSE$ profile from the training is compared against the truth in the training data in Figure~\ref{fig:oneD_channel_with_bump_UDE_result}(a). Both $WSE$ profiles are in perfect agreement. The total loss $L_{\text{total}}$ and its two components, $L_{\text{WSE}}$ and $L_{\text{uv}}$, are shown in Figure~\ref{fig:oneD_channel_with_bump_UDE_result}(b). The loss was reduced by more than four orders of magnitude. In the final iteration steps, the major loss was from the velocity mismatch $L_{\text{uv}}$ and it cannot be reduced further using the ADAM optimizer. In the future, other optimizers such as the L-BFGS-B optimizer may be used to fine-tune the loss function to further reduce the loss. \Hydrograd supports the use of a combination of optimizers in sequence. 

The good match in the WSE profiles is because the NN in the USWEs perfectly approximates the functional relationship $n(h)$ in Equation~\ref{eqn:sigmoid_n_h}. Figure~\ref{fig:oneD_channel_with_bump_UDE_result}(c) shows the comparison of Manning's $n$ values from the NN and the truth in the training data. All the data points fall on the diagonal line suggesting perfect match. Further, Figure~\ref{fig:oneD_channel_with_bump_UDE_result}(d) shows approximated functional relationship $n(h)$ by the NN and the truth (the Sigmoid function). It is observed that the NN can indeed learn the flow resistance physics  from the training data. From the figure, the most significant deviation is in the transition region of the Sigmoid function near $h=h_{\text{mid}}$. This can be attributed to at least two reasons. First, the transition region of the Sigmoid function is very narrow ($k$ has a large value of 100) and the NN may not be able to capture the transition accurately. Second, the majority of the water depth in the domain is not in the transition region. Thus, there is simply not enough data to train the NN to learn the dynamics in the transition region accurately. 

\subsubsection{The Savannah River Case with Multiple Roughness Zones for $n(h, U_{mag}, k_s)$}

\begin{figure}[htp]
    \centering
    \includegraphics[width=1\linewidth]{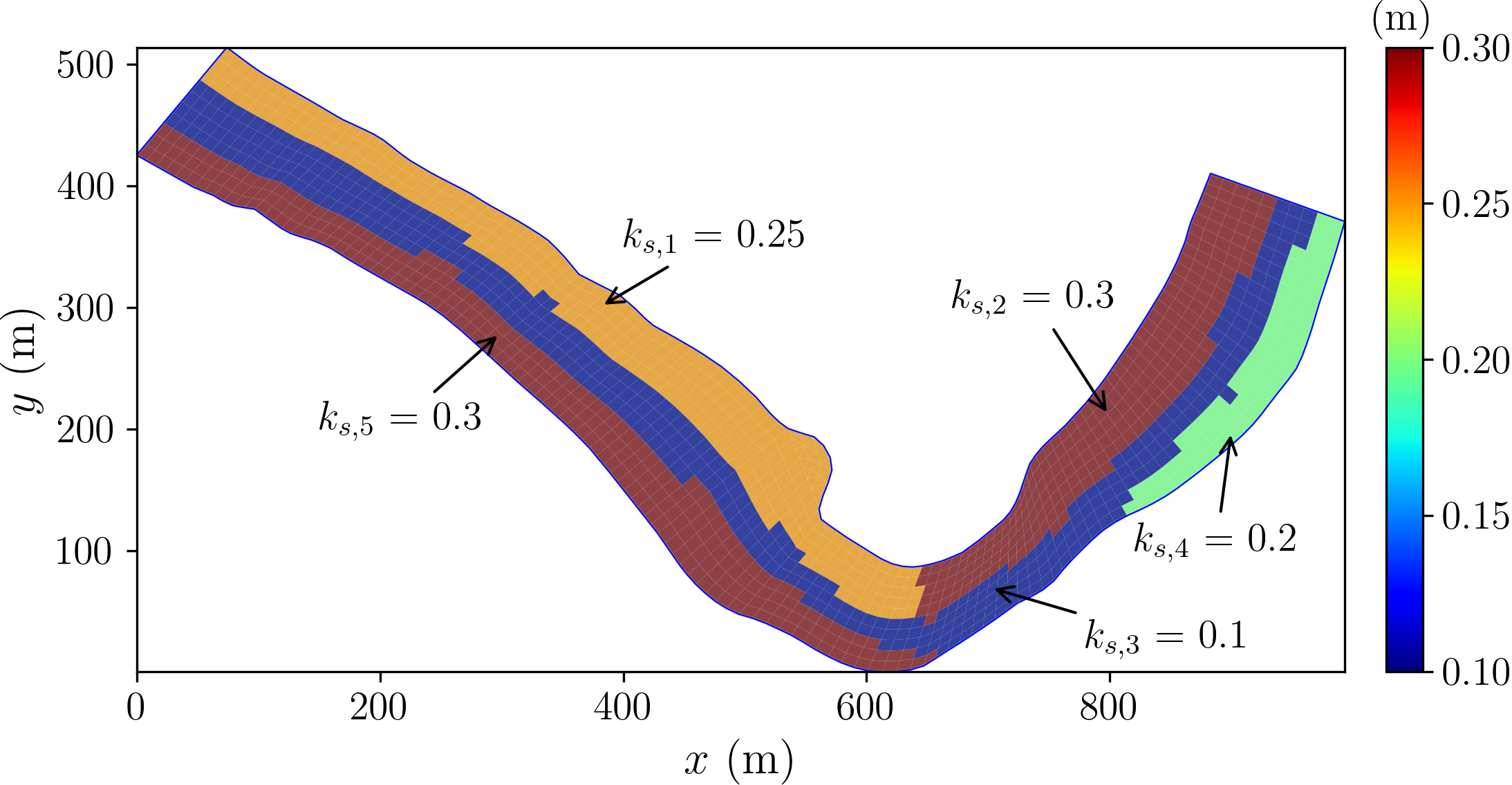}
    \caption{The contour plot of the roughness height $k_s$ for the Savannah River case.}
    \label{fig:Savana_UDE_ks_contour}
\end{figure}

This application example shows how to use the USWEs solver to recover the flow resistance law in Equation~\ref{eqn:Cheng_f_Re_hks}. This case is more complex than the 1D case above and close to real-world applications. The domain of the Savannah River case is again divided into five roughness zones. Instead of directly specifying the Manning's $n$ for each zone, the roughness height $k_s$ is specified. Figure~\ref{fig:Savana_UDE_ks_contour} shows the zoning of $k_s$ and their corresponding values. In practice, the specification of $k_s$ values is more straightforward than the Manning's $n$ because $k_s$ can be directly quantified from sediment size distribution, bathymetry, and even vegetation condition \cite{Yen2002,ChenEtAl2019}. In this case, the main channel has a smaller roughness height value in comparison with the values in the floodplain. 

Forward simulation was firstly performed to generate the training data. Specifically, during the forward simulation, the friction coefficient $f$ using Equation~\ref{eqn:Cheng_f_Re_hks} and then the Manning's $n$ were computed for each cell in the mesh at each time step. Thus, even within the same roughness zone, the friction coefficient $f$ and the Manning's $n$ are not uniform. The forward simulation was run to equilibrium and the final flow field was used for USWEs training. Both $WSE$ and flow velocity were used in the loss calculations with equal weights. The NN has three inputs and one output, i.e., it models the functional relationship $n$ = $n(h, U_{mag}, k_s)$. Instead of the dimensionless formulation in Equation~\ref{eqn:Cheng_f_Re_hks}, the NN uses the dimensional variables for the convenience of code implementation and efficiency (no need for unnecessary conversions and tracking the gradient of these operations). Both dimensionless and dimensional functional relationships are equivalent and they embed the same physical laws. The NN setup is the same as the one used in the previous section for the $n$=$n(h)$ case with the exception of the number of input variables. The training was performed for 150 iterations using the ADAM optimizer with an initial learning rate of 0.1.

\begin{figure}[htp]
    \centering
    \includegraphics[width=1\linewidth]{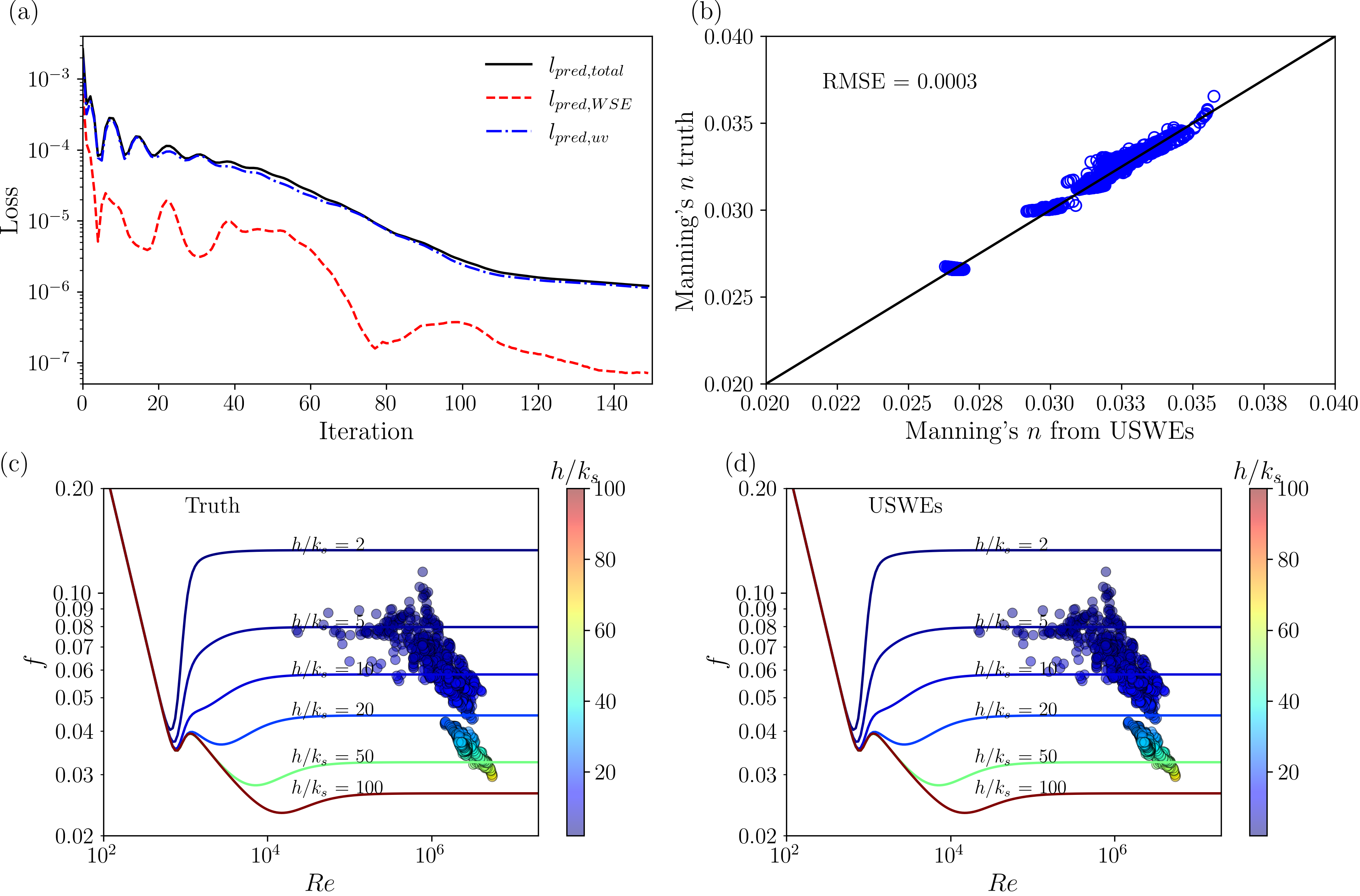}
    \caption{The training history using USWEs for the Savannah River case: (a) The history of the losses, (b) The comparison of the Manning's $n$ values from the NN at the 150th iteration and the truth. (c) The distribution of the truth friction factor $f$ for each cell in the domain on the $f(Re, h/k_s)$ plane. The lines are from Equation~\ref{eqn:Cheng_f_Re_hks} and the scatters are from all computational cells. Both the lines and the scatters are colored with $h/k_s$. (d) The distribution of the USWEs' predicted friction factor $f$ for each cell in the domain on the $f(Re, h/k_s)$ plane.}
    \label{fig:Savana_River_UDE_result_training_history_ManningN_0150}
\end{figure}

The training converged and the flow resistance was successfully recovered. Figure~\ref{fig:Savana_River_UDE_result_training_history_ManningN_0150}(a) shows the training loss history. For this case, the total loss is dominated by the velocity loss from the very beginning. The $WSE$ loss plays a relatively minor role. This is in contrast with the previous case for 1D flow over a bump, in which the $WSE$ has significant variations. In the Savannah River case, the $WSE$ has much less variations because the flow is subcritical over a relatively short reach. However, the velocity components change significantly because of the river bend. Figure~\ref{fig:Savana_River_UDE_result_training_history_ManningN_0150}(b) shows the comparison of the Manning's $n$ values from the final training step and the truth. The difference, measured by the root mean square error (RMSE) of 0.0003, is very small, indicating the accuracy of the trained neural network. To further illustrate how the flow resistance is distributed on the $f(Re, h/k_s)$ graph as proposed in \citeA{Cheng2008}, Figure~\ref{fig:Savana_River_UDE_result_training_history_ManningN_0150}(c) and (d) show the truth and the USWEs result, respectively. In these plots, the lines are from Equation~\ref{eqn:Cheng_f_Re_hks}. In the plots, each scatter point represents a computational cell in the domain. The lines and the scatters are colored with $h/k_s$. The two plots are almost identical, indicating the NN in USWEs fully recovered the flow resistance law embedded in the training data. The scatter points cover a wide range of $Re$ and $h/k_s$ within the simulation domain, showing the superb learning capabilities of the USWEs solver. 

\begin{figure}[htp]
    \centering
    \includegraphics[width=1\linewidth]{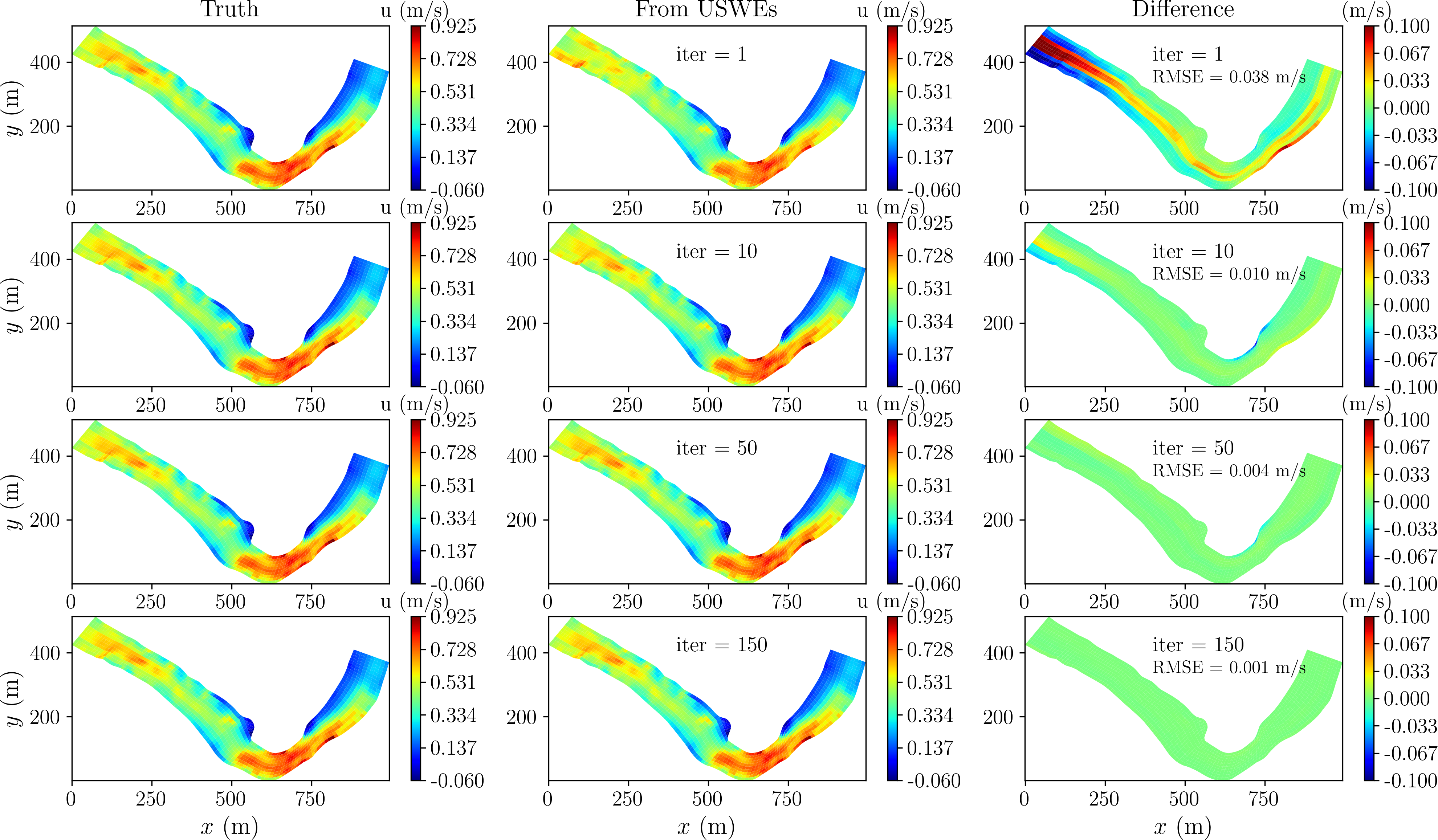}
    \caption{The comparison of the velocity component $u$ between the truth and the USWEs' prediction at four different iterations during the training process. The first column shows the truth, the second column shows the USWEs' prediction, and the third column shows the difference between the two.}
    \label{fig:Savana_UDE_U_iterations}
\end{figure}

\begin{figure}[htp]
    \centering
    \includegraphics[width=1\linewidth]{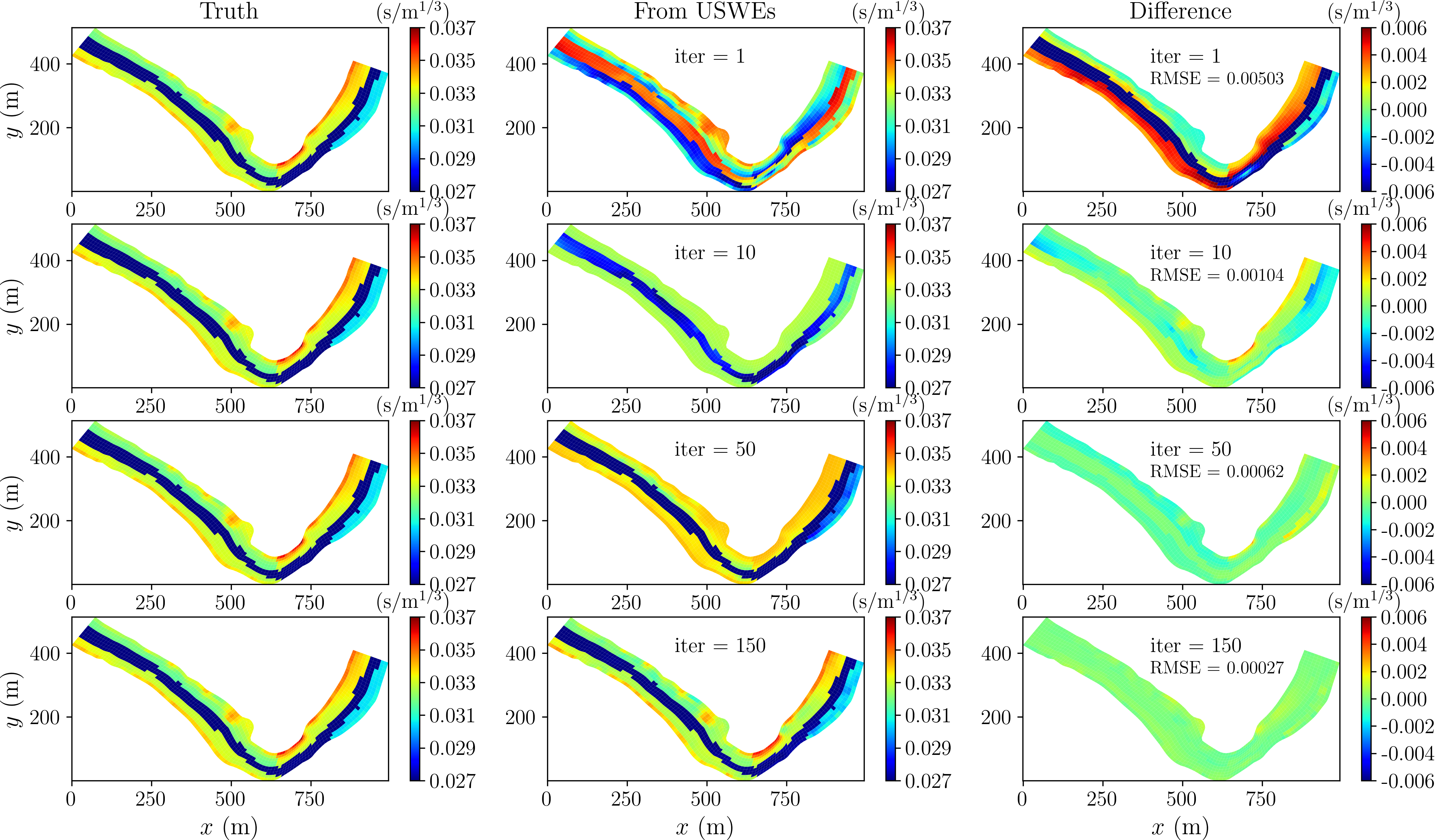}
    \caption{The comparison of the Manning's $n$ values between the truth and the USWEs' prediction at four different iterations during the training process. The first column shows the truth, the second column shows the USWEs' prediction, and the third column shows the difference between the two.}
    \label{fig:Savana_UDE_ManningN_iterations}
\end{figure}

With the USWEs solver, the training process also revealed important physical insights. Figure~\ref{fig:Savana_UDE_U_iterations} shows the comparison of the velocity component $u$ between the USWEs result and the truth at four different iterations, i.e., iteration 1, 10, 50, and 150. The first column shows the truth, the second column shows the USWEs results, and the third column shows the corresponding difference. It can be observed that the difference, measured by the RMSE, decreases rapidly with the training iterations. Similar trend can be observed from the plots for the velocity component $v$ and $WSE$. To reduce the length of the paper, these extra plots and the inversion process animation can be found in the Supplementary Information. Corresponding to Figure~\ref{fig:Savana_UDE_U_iterations}, the Manning's $n$ values from the USWEs' NN prediction and the truth are shown in Figure~\ref{fig:Savana_UDE_ManningN_iterations}. At the first iteration, the largest velocity difference is in the main channel, i.e., zone 3 with the roughness height $k_{s,3}$. This is because the Manning's $n$ in the main channel is much higher than the truth. To drive down the training loss, the optimizer quickly adjusted the NN such that the predicted main channel flow resistance was lowered (see the Manning's $n$ values at iteration 10 in Figure~\ref{fig:Savana_UDE_ManningN_iterations}). The quick reduction of loss in just a few iterations is due to the strong relationship between the flow resistance and the velocity distribution.  

\section{Discussion} \label{section:discussion}

Current scientific machine learning for flow resistance is based on synthetic cases where the flow data is simulated using empirical formulas of Manning's $n$ or flow resistance. While these methods successfully recover the prior formulas, models trained on these synthetic datasets also inherit their biases and limitations, restricting SciML’s ability to discover new physics. For example, current flow resistance formulas do not account for the temporal variation of flow resistance due to sediment transport, seasonal vegetation growth, and unsteady flow conditions. Consequently, the learned neural networks also do not take into consideration these factors. To learn a more real relationship for flow resistance, observational data for flow and ground condition should be used. Such data may not be always available at high spatial and temporal resolutions and often contains significant uncertainties due to measurement errors. USWEs provides a tool and a testbed to address these challenges in the future.  

The flow resistance relationship is learned and adapted on a case-by-case basis in this work. Is it possible to learn a more universal relationship that applies at the national and global scales? Recently, purely data-driven models have been developed to estimate Manning's $n$ at the gauge basin scale (spatially averaged over the upstream area of the gauge) across the entire USA \cite{al2024spatiotemporal}. However, these models are purely data-driven and lack spatial variation at higher resolutions. Differentiable programming also has been used to learn Manning's $n$ for each river reach at national scale using the Muskingum Cunge routing model, which is a grossly simplified model \cite{bindas2024improving,song2024high}. In such models, the momentum equation is not directly solved. Currently, there is no such large-scale differentiable hydrodynamic model which solves SWEs. In the future, USWEs in \Hydrograd needs to be expanded to support large-scale training to fill this gap. 

Despite the availability of multiple AD backends in \Hydrograd through the Julia and SciML ecosystems, we only adopted Forward-mode AD for USWEs, which uses operator overloading and dual number. Forward-mode AD is memory-efficient and well-suited for problems with few parameters (like Manning's $n$ in limited number of zones or the weights of small NNs). It becomes computationally expensive for problems in high-dimensional parameter spaces. With the forward-mode AD, the sensitivity can only be computed for one parameter at a time. One remedy for this is to use parallel computing, e.g., compute the sensitivity of multiple parameters in parallel because they are independent. To make the forward-mode AD work for large parameter numbers, future work needs to utilize the massively parallel computing power of GPUs. On the other hand, reverse-mode AD is particularly suited for problems with many parameters (like bathymetry inversion). However, both modes of AD may not be always applicable due to the high-complexity of PDE solvers. Some special treatments are needed. One such treatment is the use of the adjoint method to reduce large memory demands and mitigate gradient explosion when the PDE solver is compute-intensive and the gradient/Jacobian are needed in a large parameter space \cite{chen2018neural,onken2020discretize,lienen2022torchode,song2024ancient}. Both reverse-mode AD and adjoint-based AD are available in Julia and \Hydrograd but has only been tested for small problems. More details about AD backends in \Hydrograd can be found in the Supplementary Information.

The hydrodynamics governing equations are solved in this work using an explicit scheme with the method of line approach. The governing equations (both the continuity and momentum equations) are solved together, which makes the resulted ODE system very stiff. All these impose severe limitations on the time step size and the stability of the numerical solution. In contrast, most of popular computational hydraulics models such as SRH-2D and HEC-RAS 2D uses implicit schemes. In addition, during each time step, the coupling between the continuity and momentum equations is through an iterative algorithm. This allows for a larger time step size. In the future, the same approaches can be adopted in \Hydrograd. However, there are some challenges to be addressed. For example, it may break the current AD backends since implicit schemes requires the solution of large sparse linear equation systems, which often require iterative solvers. It is inefficient or even impossible to propagate the gradient through these linear solvers with either forward or backward-mode AD. As mentioned above, adjoint-based AD is a potential solution, which can bypass the need of gradient tracking in the iterations \cite{song2024ancient}.    

The differentiability of hydrodynamics models enables the use of gradient-based optimization algorithms for many applications which are expensive or even impossible to be solved by traditional non-differentiable models. However, it is the authors' opinion that not all hydrodynamic models must be differentiable. For many applications, non-differentiable models are sufficient. Differentiable models are computationally more expensive regardless which AD backend and approach are used. 

\section{Conclusions} \label{section:conclusions}

In this work, we proposed the Universal Shallow Water Equations, which integrates neural networks with governing partial differential equations. The USWEs are solved with a differentiable programming language and its supporting ecosystem for scientific machine learning. The developed model supports efficient automatic differentiation, enabling gradient tracking throughout the entire computational pipeline for sensitivity analysis and parameter inversion. Additionally, the connected NNs can replace poorly represented modules in the physics-based model and be trained alongside the numerical solver for knowledge discovery.

We successfully inversely modeled the spatial distribution of flow resistance in a real-world river channel and used an NN to replace the empirical formulas for flow resistance in SWEs. The embedded NN learned a general relationship for flow resistance, a task that has been highly challenging for traditional models. This approach paves the way for learning more complex functional relationships between Manning’s $n$ and its influencing factors using observational data, as well as addressing other inversion problems and knowledge discovery in hydrodynamic modeling.

\section*{Data Availability Statement}
A permanent copy of code, data, and scripts used for this work has also been achieved in the CUAHSI's
HydroShare: 

\url{http://www.hydroshare.org/resource/c8ea7e80be434a228038d1f613a9b403}. 

The code and cases are also publicly available at \url{https://github.com/psu-efd/Hydrograd.jl}.

\acknowledgments
This work was supported by a seed grant from the Institute of Computational and Data Sciences at the Pennsylvania State University. We also greatly benefited from the discussions with Dr. Chaopeng Shen on the topic of differentiable programming and its use in computational modeling.

\bibliography{references}

\end{document}